# $M$C$_6$Li$_6$ ($M$ = Li, Na and K): A New Series of Aromatic Superalkalis

Ambrish Kumar Srivastava

*Department of Physics, Deen Dayal Upadhyaya Gorakhpur University, Civil Lines, Gorakhpur 273009, India*

E-mail: ambrishphysics@gmail.com;  aks.ddugu@gmail.com

**Abstract**

Organic superalkalis are carbon-based species possessing lower ionization energy than alkali atom. In the quest for new organic superalkalis, we study the $MC_6Li_6$ ($M$ = Li, Na, and K) complexes and their cations by decorating hexalithiobenzene with an alkali atom using density functional theory. All $MC_6Li_6$ complexes are planar and stable against dissociation into $M$ + $C_6Li_6$ fragments, irrespective of their charge. These complexes are stabilized by charge transfer from $M$ to $C_6Li_6$, although the back-donation of charges tends to destabilize neutral species. Furthermore, their degree of aromaticity increases monotonically from $M$ = Li to K, unlike $MC_6Li_6^+$ cations, which are not aromatic as suggested by their NICS values. The adiabatic ionization energies of $MC_6Li_6$ (3.08-3.22 eV) and vertical electron affinities of $MC_6Li_6^+$ (3.04-3.15 eV) suggest that $MC_6Li_6$ species form a new series of aromatic superalkalis. The variation of the ionization energy of $MC_6Li_6$ is found to be in accordance with the NICS values of $MC_6Li_6^+$. The superalkali nature of $MC_6Li_6$ and its relation with NICS values are explained on the basis of the positive charge delocalization. We believe that these species will not only enrich the aromatic superalkalis but also their possible applications will be explored.

## 1. Introduction

Hexalithiobenzene ($C_6Li_6$) is an organo-metallic analogue of benzene, a prototype aromatic hydrocarbon. Shimp *et al.* [1] synthesized $C_6Li_6$ molecule in 1978, which was isolated by Baran *et al.* [2] in 1992. It has a beautiful star-like structure complemented by the binding of Li to carbon skeletons, which has attracted the attention of several groups [3-6]. Moreno *et al.* [7], however, described the lowest-energy structure as three $C_2^{2-}$ fragments strongly aggregated through lithium bridges. Apart from its interesting structural features, the $C_6Li_6$ molecule has been employed for several applications. For instance, Raptis *et al.* [8] recognized the exceptionally high second-order hyperpolarizability of $C_6Li_6$, which is responsible for the third harmonic generation. Furthermore, the applications of $C_6Li_6$ in hydrogen storage [9, 10] and $CO_2$-storage [11] have been reported. In a recent study, the low ionization energy (IE) feature of $C_6Li_6$ has been highlighted [12] and it was noticed that the vertical IE of $C_6Li_6$ (4.48 eV) is lower than the IE of Li atom (5.39 eV) [13].

The species, having lower IE than an alkali atom, are referred to as superalkalis [14]. According to Gutsev and Boldyrev [14], the typical superalkalis belong to the formula of $XM_{k+1}$, where $k$ is the nominal valence of electronegative core $X$ and $M$ is an alkali metal. These species are very interesting not only due to their unusual structures but also due to the distinguished properties of their compounds. Due to their lower IE, superalkalis possess strong reducing properties and can reduce $CO_2$, $NO$, $NO_2$, *etc.* [15-18]. Nevertheless, superalkalis can form a variety of charge transfer compounds including supersalts [19-22], superbases [23-25], alkalides [26-28], *etc.* The superalkali behaviour is not only confined to inorganic species. There exist some organic superalkalis as well including $CO_3Li_3^+$ superalkali cation [29], polynuclear superalkali cations based on alkali-monocyclic (pseudo)-oxocarbon [30], superalkali cations based on $C_4H_{24}^-$ and $C_5H_5^-$ aromatic anions [31]. These all species are organo-metallic due to the presence of alkali atoms. Superalkalis can also be designed without alkali atoms. Hou *et al.* [32] proposed $C_2H_9^+$ as the first non-metallic organic superalkali cation. Subsequently, other non-metallic organic superalkali cations such as $C_5NH_6^+$ and $P_7R_4^+$ [R = Me, CH$_2$Me, CH(Me)$_2$ and C(Me)$_3$] have been reported by Zhao *et al.* [15] and Giri *et al.* [33], respectively. Recently, we have reported $C_xH_{4x+1}^+$ species (x =1-5) as a series of non-metallic organic superalkali cations [34]. Although much progress has been made in this field [35], superalkalis have been a subject of continuous investigations due to their ever-expanding applications. In this paper, we report a new class of organic, planar and aromatic superalkalis $MC_6Li_6$ by the interaction of alkali atom $M$ with $C_6Li_6$ for $M$ = Li, Na and K using density functional theory (DFT) based method.

2. **Computational details**

All DFT calculations in this study were performed at the B3LYP method [36, 37] using a 6-311++G(d) basis set in Gaussian 09 program [38]. This scheme has been already employed in our previous study on $C_6Li_6$ [12]. The geometry optimization was carried out without any symmetry constraints and followed by frequency calculations to ensure that the optimized structures correspond to true minima in the potential energy surface. The binding energies ($E_b$'s) of neutral and cationic $MC_6Li_6$ complexes are calculated as below:

$$E_b = E[M^{0,+}] + E[C_6Li_6] - E[MC_6Li_6^{0,+}]$$

where $E[..]$ represents the total electronic energy including zero-point correction and superscripts indicate the charge of respective species. The adiabatic ionization energy (IE$_a$) of $MC_6Li_6$ is calculated by the difference of total energy of equilibrium neutral and cationic structures. The vertical electron affinity (EA$_v$) of $MC_6Li_6^+$ is obtained by difference of total energy of equilibrium cation and single-point energy of corresponding neutral at the equilibrium geometry of cation:

IE$_a$ = $E_{cation}[MC_6Li_6^+]$ − $E_{neutral}[MC_6Li_6]$

EA$_v$ = $E_{cation}[MC_6Li_6^+]$ − $E_{neutral}[MC_6Li_6^+]$

## 3. Results and discussion

The structure of the planar $C_6Li_6$ is displayed in Fig. 1. The ring bond lengths ($d_{C-C}$) of $C_6Li_6$ are 1.420 Å and C-Li bond lengths ($d_{C-Li}$) are 1.914 Å as listed in Table 1. The equilibrium structures of $MC_6Li_6$ complexes ($M$ = Li, Na and K) are shown in Fig. 2. One can see that $M$ interacts with the $C_6Li_6$ ring such that the ring remains almost planar with the ring-distortion angle of less than $0.05^0$. The distances between the center of the ring and $M$ ($d_{ring-M}$) increase from 1.605 Å for $M$ = Li to 2.645 Å for $M$ = K. The $d_{C-C}$ and $d_{C-Li}$ of $MC_6Li_6$ lie in the range 1.424-1.432 Å and 1.932-1.945 Å, respectively. Thus, there is only a marginal change (< 0.05 Å) in $d_{C-C}$ and $d_{C-Li}$ of $MC_6Li_6$ as compared to those of $C_6Li_6$. Fig. 2 also displays the equilibrium structures of $MC_6Li_6^+$ cations. Like their neutral counterparts, the ring-distortion is almost negligible (< $0.03^0$) such that the ring remains planar and bond-lengths ($d_{C-C}$) range 1.430-1.437 Å. However, Li atoms in $MC_6Li_6^+$, unlike $MC_6Li_6$, go slightly out-of-plane as clearly visible in the side-view of Fig. 2. The bond-length $d_{C-Li}$ of the cation is slightly increased as compared to $MC_6Li_6$, except for $M$ = Li. On the contrary, the $d_{ring-M}$ of $MC_6Li_6^+$ is slightly decreased as compared to their neutral species for $M$ = Na and K but not for $M$ = Li.

$C_6Li_6$ is a lithiated analogue of benzene, a prototype aromatic molecule. However, the degree of aromaticity of $C_6Li_6$ is reduced as compared to benzene due to Li-substitution, i.e., $\sigma$-electron contribution [12]. Therefore, it is interesting to observe the aromaticity of $MC_6Li_6$, which is quantified by calculating the nucleus independent chemical shift (NICS) at the center of the ring. NICS is the most popular criterion of aromaticity [39, 40], based on the ring current due to the induced magnetic field. $NICS_{zz}$, a component of NICS along the direction perpendicular to the ring-plane, measures the contribution of $\pi$-electrons in the ring current. The positive value of NICS indicates anti-aromaticity whereas negative values advocate aromaticity. Although the $NICS_{zz}$ value of $C_6Li_6$ is -25.11 ppm, its NICS value is only -3.38 ppm. The negative NICS value increases monotonically in the case of $M$C$_6$Li$_6$, ranging between -3.97 ppm for $M$ = Li and -5.00 ppm $M$ = K. However, the NICS$_{zz}$ value of $M$C$_6$Li$_6$ varies between -23.65 ppm and -27.52 ppm. The larger (negative) NICS values of $M$C$_6$Li$_6$ as compared to C$_6$Li$_6$ may suggest an increase in the degree of aromaticity. To compare, the NICS and NICS$_{zz}$ values of aromatic benzene are -8.05 ppm and -14.49 ppm, respectively at the same level. To explain this, we have plotted electron localization function (ELF) maps of $M$C$_6$Li$_6$ in Fig. 3. The increase in aromaticity is evidently due to the increase in the delocalization of $\pi$-electron density in the ring. It is also interesting to note that the magnitude of NICS is significantly reduced in the case of $M$C$_6$Li$_6^+$. For instance, it becomes -1.21 ppm for $M$ = K and even positive (+0.55 ppm) for $M$ = Na. This is due to the contribution of $\sigma$-electrons, which supports paratropic ring currents and reduces the aromaticity significantly. In the discussion below, we will show that the NICS values of $M$C$_6$Li$_6^+$ are closely related to the superalkali behavior of $M$C$_6$Li$_6$ species.

The stability of neutral and cationic $M$C$_6$Li$_6$ complexes has been verified by calculating their binding energy ($E_b$) against dissociation into $M$ + C$_6$Li$_6$ and $M^+$ + C$_6$Li$_6$ fragments, respectively. The calculated $E_b$ values are found to be positive as listed in Table 1 and hence, all $M$C$_6$Li$_6$ complexes are stable. Furthermore, $M$C$_6$Li$_6^+$ cations are relatively more stable than their neutral counterparts due to higher $E_b$ values. However, the stability of $M$C$_6$Li$_6^+$ cations decreases from $M$ = Li to K due to the increase in their distance from the ring (see Table 1). Since $M$C$_6$Li$_6$ complexes are stabilized by the charge-transfer between $M$ atom and C$_6$Li$_6$ moiety. Therefore, $E_b$ measures the strength of this charge-transfer interaction. We will later show that there is some back-donation of charges, which will eventually explain the trend of $E_b$ values.

The superalkali behavior of $M$C$_6$Li$_6$ complexes has been explored by their IE$_a$ values and EA$_v$ of corresponding cations. EA$_v$ of cations is equivalent to IE$_a$ of their neutral counterparts, except the fact that the former does not account for geometric relaxation. One can note that $IE_a$ of $MC_6Li_6$ lies in the range 3.08-3.22 eV whereas $EA_v$ of $MC_6Li_6^+$ ranges 3.04-3.15 eV. Evidently, these values are small as compared to the IE of alkali metal, whose maximal value is limited to 5.39 eV for Li [13]. Therefore, $MC_6Li_6$ complexes behave as superalkalis. In a previous study, the vertical IE of $C_6Li_6$ is reported to be 4.48 eV [12], which is smaller than the IE of Na (5.14 eV) and even comparable to that of K (4.34 eV). Therefore, the charge transfer takes place from $C_6Li_6$ to Li in the $LiC_6Li_6$ complex. On the contrary, the charge is transferred from Na and K to $C_6Li_6$ in $NaC_6Li_6$ and $KC_6Li_6$ complexes, respectively. In addition, there is a back-donation of charges from $C_6Li_6$ to Na and K. This is reflected in the highest occupied molecular orbitals (HOMOs) of $MC_6Li_6$ complexes plotted in Fig. 4. One can see that the HOMO of $LiC_6Li_6$ is contributed by peripheral Li atoms of $C_6Li_6$ moiety. The HOMOs of $NaC_6Li_6$ and $KC_6Li_6$ are mainly composed of Na and K atoms, respectively with some contribution from Li atoms. The NBO charges, listed in Table 2, also support the back-donation of charges. For instance, the charge on $M$ ($Q_M$) in $MC_6Li_6$ complexes is +0.60e for $M$ = Li, which is significantly reduced to +0.28e and +0.46e for $M$ = Na and K, respectively. This back-donation of charges tends to destabilize the $MC_6Li_6$ complexes ($M$ = Na and K), decreasing their binding energies as compared to $LiC_6Li_6$ (see Table 1).

Although all $MC_6Li_6$ complexes belong to the class of superalkalis, the $IE_a$ of $NaC_6Li_6$ is larger than that of $LiC_6Li_6$. The trend of $IE_a$ values can be explained on the basis of NBO charge distribution on $MC_6Li_6$ complexes and $MC_6Li_6^+$ cations (as listed in Table 2). Note that traditional superalkalis employ a central (electronegative) atom attached with electropositive ligands and their low IE results due to an increase in the charge delocalization over electropositive ligands. One can see that the charges on C atoms ($Q_C$) in $MC_6Li_6$ and $MC_6Li_6^+$ differ only slightly. In $LiC_6Li_6^+$, the positive charge is almost equally distributed over all Li atoms. In the case of $NaC_6Li_6^+$, most of the positive charge (58%) is localized on Na atom as $Q_{Na}$ increases to +0.86e from +0.28e, which results in higher $IE_a$ of $NaC_6Li_6$ complex. For $KC_6Li_6^+$, the localization of positive charge decreases to 43%, which tends to reduce the $IE_a$ of $KC_6Li_6$. Thus, the lower $IE_a$ of $KC_6Li_6$ results due to an increase in the delocalization of positive charges over Li atoms.

Having established that $MC_6Li_6$ complexes are aromatic and they behave as superalkalis, we might be interested in whether there exists any relation between NICS and $IE_a$ values of these species. In Fig. 5, we have plotted the NICS of $MC_6Li_6^+$ and $IE_a$ of $MC_6Li_6$ complexes (or equivalently $EA_v$ of $MC_6Li_6^+$, see Table 2). We notice that the variation of NICS follows the same trend as that of $IE_a$. This can be expected due to the fact that lower $IE_a$ results due to the delocalization of charges, which supports diatropic current and results in negative NICS values. On the contrary, the localization of charge causes the increase in $IE_a$, which leads to positive NICS value (as in the case of $NaC_6Li_6$). Thus, both these properties are associated with the delocalization of electrons (charges) and consequently, seem to be related to each other.

## 4. Conclusions

Using a density functional approach, we have studied the interaction of an alkali atom ($M$) with $C_6Li_6$ for $M$ = Li, Na, and K. The resulting $MC_6Li_6$ complexes have been explored in their neutral as well as cationic forms. We have noticed both neutral and cationic $MC_6Li_6$ structures resemble each other. In addition to the charge transfer from $M$ to $C_6Li_6$, there is a back-donation of charges in neutral species. This tends to destabilize neutral species, which is reflected in their lower binding energies as compared to their cations. The NICS values and ELF maps suggest that the degree of aromaticity of $MC_6Li_6$ increases monotonically from $M$ = Li to K, unlike $MC_6Li_6^+$ cations, which are not aromatic. The $IE_a$ of $MC_6Li_6$ (3.08-3.22 eV) and $EA_v$ of $MC_6Li_6^+$ (3.04-3.15 eV) are small enough to suggest the superalkali nature of $M$C$_6$Li$_6$ species. We have also noticed that the variation of IE$_a$ of $M$C$_6$Li$_6$ is consistent with that of NICS values of $M$C$_6$Li$_6^+$. This surprising relation can be explained on the basis of the fact that both superalkali property and aromaticity are associated with the delocalization of charges. We believe that these findings will get further attention from the researchers worldwide. Further applications of these aromatic superalkalis are in progress in our lab and shall be reported shortly.

**Acknowledgement**

Dr. A. K. Srivastava acknowledges Prof. N. Misra, Department of Physics, University of Lucknow, Prof. S. N. Tiwari, Department of Physics, DDU Gorakhpur University, for helpful discussions and University Grants Commission (UGC), New Delhi, India for approving Start Up project [Grant No. 30-466/2019(BSR)].

**References**

[1] L. A. Shimp, C. Chung and R. J. Lagow, Inorg. Chim. Acta 29 (1978) 77.

[2] R. J. Baran, Jr., D. A. Hendrickson, D. A. Laude, Jr. and R. J. Lagow, J. Org. Chem. 57 (1992) 3759.

[3] Y. Xie, H. F. Schaefer III, Chem. Phys. Lett. 179 (1991) 563.

[4] B. J. Smith, Chem. Phys. Lett. 207 (1993) 403.

[5] S. M. Bachrach, J. V. Miller Jr., J. Org. Chem. 67 (2002) 7389.

[6] Y.-B. Wu, J.-L. Jiang, R.-W. Zhang, Z.-X. Wang, Chem. Eur. J. 16 (2010) 1271.

[7] D. Moreno, G. Martinez-Guajardo, A. Diaz-Celaya, J. M. Mercero, R. de Coss, N. Perez-Peralta, G. Merino, Chem. Eur. J. 19 (2013) 12668.

[8] S. G. Raptis, M. G. Papadopoulos and A. J. Sadlej, Phys. Chem. Chem. Phys. 2 (2000) 3393-3399.

[9] A. Vasquez-Espinal, R. Pino-Rios, P. Fuentealba, W. Orellana, W. Tiznado, Int. J. Hydrogen Energy 41 (2016) 5709.

[10] S. Giri, F. Lund, A. S. Nunez, A. Toro-Labbe, J. Phys. Chem. C 117 (2013) 5544.

[11] A. K. Srivastava, Int. J. Quantum Chem. 119 (2019) e25904.

[12] A. K. Srivastava, Mol. Phys. 116 (2018) 1642-1649.

[13] ] J. E. Sansonetti, W. C. Martin, S. L. Young, J. Phys. Chem. Ref. Data 34 (2005) 1559.

[14] G. L. Gutsev, A. I. Boldyrev, Chem. Phys. Lett. 92 (1982) 262–266.

[15] T. Zhao, Q. Wang, P. Jena, Nanoscale 9 (2017) 4891.

[16] A. K. Srivastava, Int. J. Quantum Chem. 119 (2018) e25598.

[17] H. Park, G. Meloni, Dalton Trans. 6 (2017) 11942.

[18] A. K. Srivastava, Chem. Phys. Lett. 695 (2018) 205–210.

[19] Y. Li, D. Wu, Z.-R. Li, Inorg. Chem. 47 (2008) 9773-9778.

[20] H. Yang, Y. Li, D. Wu, Z.-R. Li, Int. J. Quantum Chem. 112 (2012) 770-778.

[21] A. K. Srivastava, N. Misra, Mol. Phys. 112 (2014) 2621-2626.

[22] S. Giri, S. Bahera, P. Jena, J. Phys. Chem. A 118 (2014) 638-645.

[23] A. K. Srivastava, N. Misra, New J. Chem. 39 (2015) 6787-6790.

[24] A. K. Srivastava, N. Misra, Chem. Phys. Lett. 648 (2016) 152-155.

[25] M. Winfough, G. Meloni, Dalton Trans. 47 (2017) 159.

[26] W. Chen, Z.-R. Li, D. Wu, Y. Li, C.-C. Sun, J. Phys. Chem. A 109 (2005) 2920-2924.

[27] W. M. Sun, L. T. Fan, Y. Li, J. Y. Liu, D. Wu, Z. R. Li, Inorg. Chem. 53 (2014) 6170.

[28] A. K. Srivastava, N. Misra, Chem. Phys. Lett. 639 (2015) 307-309.

[29] J. Tong, Y. Li, D. Wu, Z. -J. Wu, Inorg. Chem. 51 (2012) 6081−6088.

[30] J. Tong, Z. Wu, Y. Li, D. Wu, Dalton Trans. 42 (2013) 577-584.

[31] W. M. Sun, Y. Li, D. Wu, Z. -R. Li, J. Phys. Chem. C 117 (2013) 24618−24624.

[32] N. Hou, Y. Li, D. Wu, Z. -R. Li, Chem. Phys. Lett. 575 (2013) 32–35.

[33] S. Giri, G. N. Reddy, P. Jena, J. Phys. Chem. Lett. 7 (2016) 800-805.

[34] W. M. Sun, D. Wu, Chem. Eur. J. 25 (2019) 9568-9579.

[35] A. K. Srivastava, Mol. Phys. 117 (2019) e1615648.

[36] A. D. Becke, Phys. Rev. A 38 (1988) 3098.

[37] C. Lee, W. Yang, R.G. Parr, Phys. Rev. B 37 (1988) 785-789.

[38] M.J. Frisch, G.W. Trucks, H.B. Schlegel, G.E. Scuseria, M.A. Robb, J.R. Cheeseman, G. Scalmani, V. Barone, B. Mennucci, G.A. Petersson, H. Nakatsuji, M. Caricato, X. Li, H.P. Hratchian, A.F. Izmaylov, J. Bloino, G. Zheng, J.L. Sonnenberg, M. Hada, M. Ehara, K. Toyota, R. Fukuda, J. Hasegawa, M. Ishida, T. Nakajima, Y. Honda, O. Kitao, H. Nakai, T. Vreven, J. Montgomery, J. A., J.E. Peralta, F. Ogliaro, M. Bearpark, J.J. Heyd, E. Brothers, K.N. Kudin, V.N. Staroverov, R. Kobayashi, J. Normand, K. Raghavachari, A. Rendell, J.C. Burant, S.S. Iyengar, J. Tomasi, M. Cossi, N. Rega, J.M. Millam, M. Klene, J.E. Knox, J.B. Cross, V. Bakken, C. Adamo, J. Jaramillo, R. Gomperts, R.E. Stratmann, O. Yazyev, A.J. Austin, R. Cammi, C. Pomelli, J.W. Ochterski, R.L. Martin, K. Morokuma, V.G. Zakrzewski, G.A. Voth, P. Salvador, J.J. Dannenberg, S. Dapprich, A.D. Daniels, O. Farkas, J.B.

Foresman, J.V. Ortiz, J. Cioslowski, D.J. Fox, in, Gaussian 09, Revision D. 01, Gaussian, Inc., Wallingford CT, 2009.

[39] P. v. R. Schleyer, H. Jiao, Pure Appl. Chem. 68 (1996) 209.

[40] P. v. R. Schleyer, H. Jiao, N. J. v. E. Hommes, V. G. Malkin, O. L. Malkina, J. Am. Chem. Soc. 119 (1997) 12669.

Table 1. B3LYP/6-311+G(d) calculated structural parameters, NICS and NICS$_{zz}$ of neutral and cationic $M$C$_6$Li$_6$ complexes for $M$ = Li, Na and K.

| $M$ | $M$C$_6$Li$_6$ | | | | | $M$C$_6$Li$_6^+$ | | | | |
|---|---|---|---|---|---|---|---|---|---|---|
| | $d_{C-C}$ | $d_{C-Li}$ | $d_{ring-M}$ | NICS | NICS$_{zz}$ | $d_{C-C}$ | $d_{C-Li}$ | $d_{ring-M}$ | NICS | NICS$_{zz}$ |
| | Å | Å | Å | ppm | ppm | Å | Å | Å | ppm | ppm |
| Li | 1.432 | 1.945 | 1.605 | -3.97 | -23.65 | 1.437 | 1.943 | 1.607 | -1.77 | -18.85 |
| Na | 1.426 | 1.932 | 2.257 | -4.20 | -25.47 | 1.433 | 1.939 | 2.154 | +0.55 | -20.70 |
| K | 1.424 | 1.933 | 2.645 | -5.00 | -27.52 | 1.430 | 1.936 | 2.505 | -1.21 | -22.68 |

Table 2. B3LYP/6-311+G(d) calculated NBO charges and binding energy of neutral and cationic $M$C$_6$Li$_6$ complexes for $M$ = Li, Na and K. The adiabatic ionization energy of $M$C$_6$Li$_6$ and vertical electron affinity of $M$C$_6$Li$_6^+$ are also listed.

| $M$ | $M$C$_6$Li$_6$ | | | | | $M$C$_6$Li$_6^+$ | | | | |
|---|---|---|---|---|---|---|---|---|---|---|
| | $Q_C$ | $Q_{Li}$ | $Q_M$ | $E_b$ | IE$_a$ | $Q_C$ | $Q_{Li}$ | $Q_M$ | $E_b$ | EA$_v$ |
| | e | e | e | eV | eV | e | e | e | eV | eV |
| Li | -0.65 | +0.55 | +0.60 | 1.82 | 3.14 | -0.68 | +0.72 | +0.74 | 4.30 | 3.13 |
| Na | -0.65 | +0.60 | +0.28 | 0.91 | 3.22 | -0.69 | +0.71 | +0.86 | 3.11 | 3.15 |
| K | -0.65 | +0.57 | +0.46 | 1.00 | 3.08 | -0.68 | +0.69 | +0.91 | 2.41 | 3.04 |

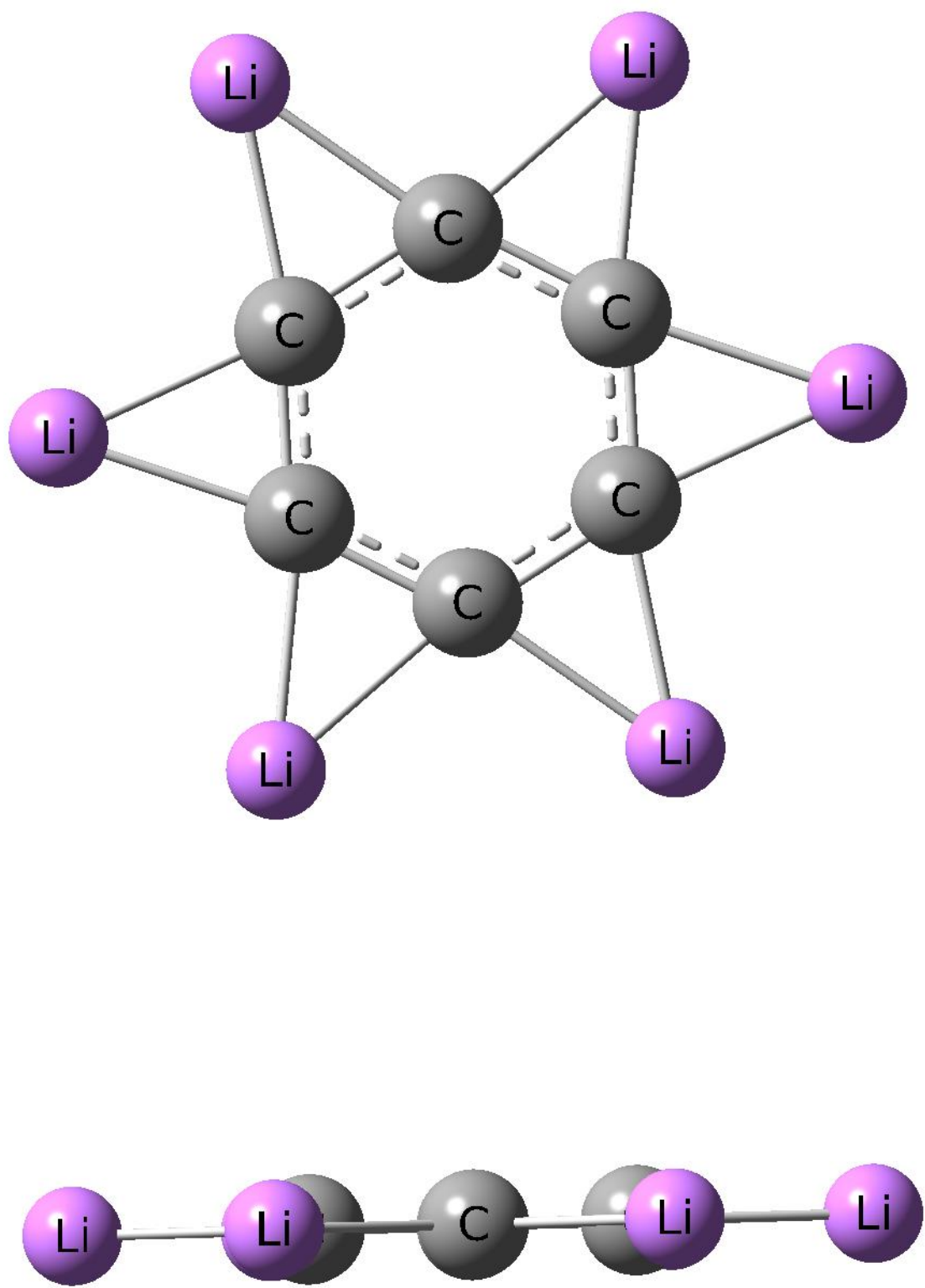

Fig. 1. The equilibrium structure of $C_6Li_6$ obtained at B3LYP/6-311+G(d) level.

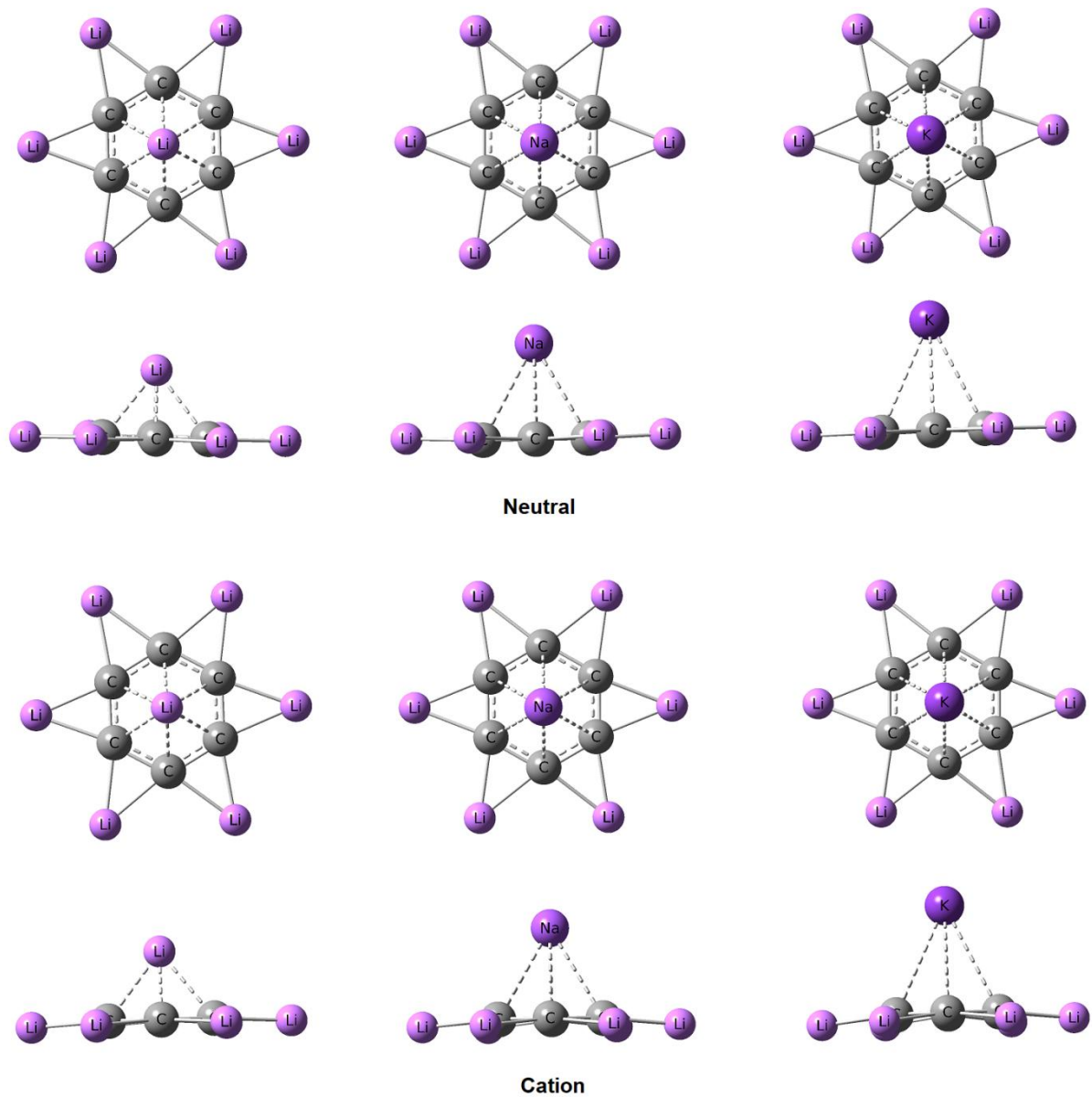

Fig. 2. The equilibrium structures of neutral and cationic $MC_6Li_6$ complexes for $M$ = Li, Na and K obtained at B3LYP/6-311+G(d) level. Front and side views are displayed.

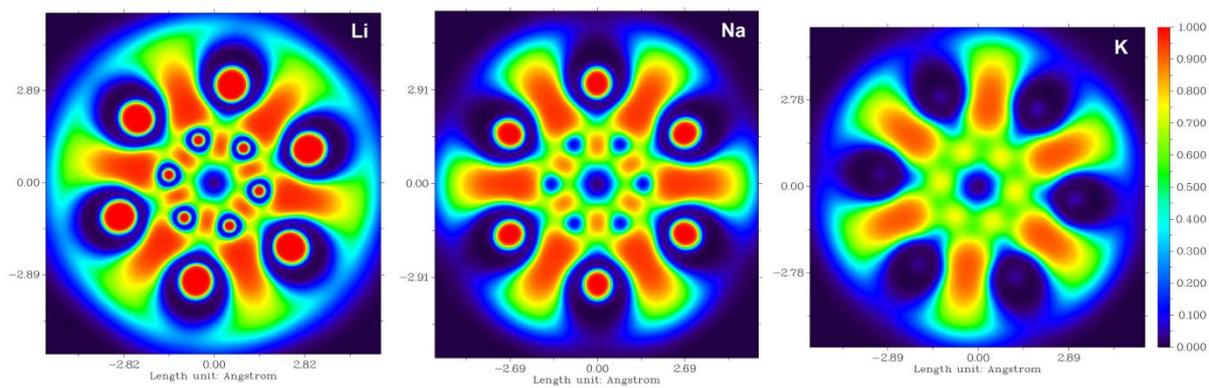

Fig. 3. The electron localization function (ELF) maps of $MC_6Li_6$ complexes for $M$ = Li, Na and K.

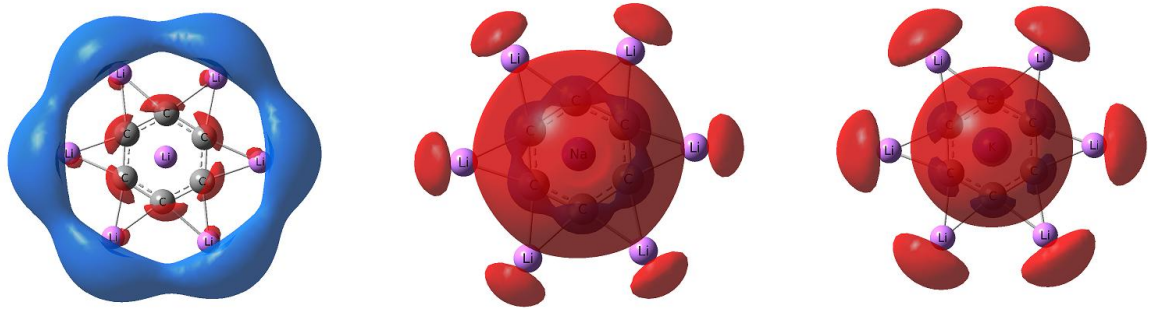

Fig. 4. The highest occupied molecular orbital (HOMO) of $M$C$_6$Li$_6$ complexes for $M$ = Li, Na and K.

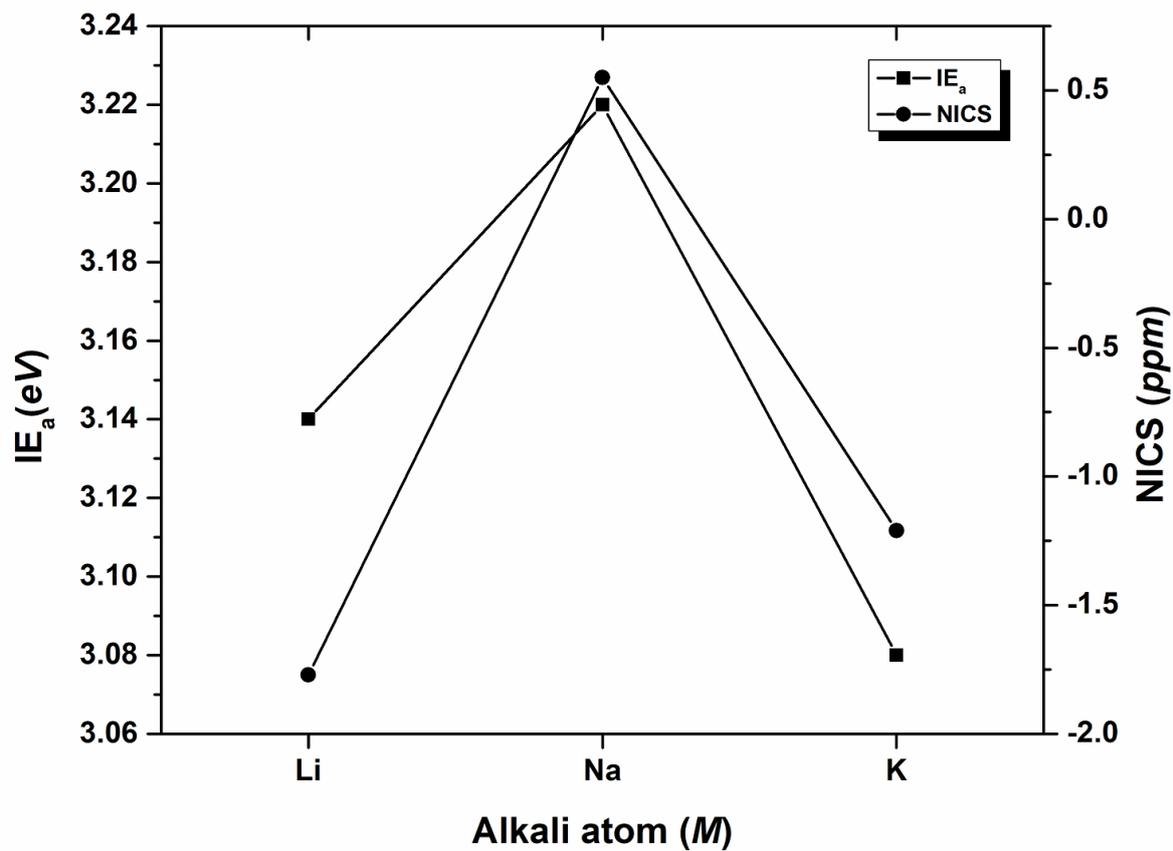

Fig. 5. The variation of adiabatic ionization energy (IE$_a$) of $M$C$_6$Li$_6$ complexes and nucleus independent chemical shift (NICS) of $M$C$_6$Li$_6^+$ cations.